\documentclass[prl,twocolumn,aps,amsmath,amssymb,longbibliography]{revtex4-1}

\usepackage{graphicx}
\usepackage{dcolumn}
\usepackage{bm}
\newcommand{\etal}{\emph{et al.}}
\newcommand{\be}{\begin{equation}}
\newcommand{\ee}{\end{equation}}
\newcommand{\bfig}{\begin{figure}}
\newcommand{\efig}{\end{figure}}

\begin{document}      
\title{Anomalous Nernst Effect in the Dirac Semimetal Cd$_3$As$_2$} 
 
\author{Tian Liang$^{1\ast,\dagger}$}
\author{Jingjing Lin$^{1\ast}$}
\author{Quinn Gibson$^{2}$}
\author{Tong Gao$^1$}
\author{Max Hirschberger$^1$}
\author{Minhao Liu$^1$}
\author{R. J. Cava$^2$}
\author{N. P. Ong$^{1}$}
\affiliation{
Departments of Physics$^1$ and Chemistry$^2$, Princeton University, Princeton, NJ 08544
} 

\begin{abstract}
{Dirac and Weyl semimetals display a host of novel properties. In Cd$_3$As$_2$, the Dirac nodes lead to a protection mechanism that strongly suppresses backscattering in zero magnetic field, resulting in ultrahigh mobility ($\sim$ 10$^7$ cm$^2$ V$^{-1}$ s$^{-1}$). In applied magnetic field, an anomalous Nernst effect is predicted to arise from the Berry curvature associated with the Weyl nodes. We report observation of a large anomalous Nernst effect in Cd$_3$As$_2$. Both the anomalous Nernst signal and transport relaxation time $\tau_{tr}$ begin to increase rapidly at $\sim$ 50 K. This suggests a close relation between the protection mechanism and the anomalous Nernst effect. In a field, the quantum oscillations of bulk states display a beating effect, suggesting that the Dirac nodes split into Weyl states, allowing the Berry curvature to be observed as an anomalous Nernst effect.
}
\end{abstract}
 
\maketitle
      
The field of topological quantum materials has recently expanded to include the Dirac (and Weyl) semimetals, which feature 3D bulk Dirac states with nodes that are protected by symmetry~\cite{Ashvin,Kane,Ran,Bernevig}. In Dirac semimetals, each Dirac cone is the superposition of two Weyl nodes which have opposite chiralities ($\chi = \pm 1$). The Weyl nodes are prevented from hybridizing by the combination of point group symmetry, inversion symmetry and time-reversal symmetry (TRS)~\cite{Bernevig}. In the presence of a magnetic field $\bf B$, the breaking of TRS leads to separation of the Weyl nodes and the appearance of a Berry curvature $\mathbf{\Omega}(\bf k)$. Because $\mathbf{\Omega}(\bf k)$ acts like an intense magnetic field, it exerts a strong force on charge carriers~\cite{Murakami2007,Murakami2008}. The first examples of Dirac semimetals, Na$_3$Bi and Cd$_3$As$_2$, were identified by Wang \etal~\cite{XiDai2012,XiDai2013}. (In the Weyl semimetal TaAs, the Weyl nodes are already well separated in zero $\bf B$ because its space group lacks inversion symmetry. The signature surface Fermi arcs were recently observed by angle-resolved photoemission experiments on TaAs~\cite{Hasan2015, YLChen2015, HongDing2015}. Surface modes in Cd$_3$As$_2$ have also been observed by Shubnikov de Haas (SdH) oscillations ~\cite{Moll2016}.)

An interesting phenomenon in Dirac and Weyl semimetals is the chiral anomaly which refers to the axial current that results from ``pumping'' electrons between left- and right-moving Dirac branches (of opposite $\chi$) when an electric field $\bf E$ is applied $\parallel\bf B$~\cite{Adler,Bell,NielsenNinomiya,SonSpivak}. Recently, the chiral anomaly was successfully observed as a large, negative longitudinal magnetoresistance (LMR) in Na$_3$Bi~\cite{Xiong2015} and GdPtBi~\cite{Max2016}. The anomaly engenders a 4- to 6-fold decrease in the longitudinal resistance in a moderate $B$. Negative LMRs have also been reported in Bi$_{1-x}$Sb$_x$~\cite{Kim2013}, Cd$_3$As$_2$~\cite{Liang2015,Xiu2015}, ZrTe$_5$~\cite{Li2015}, TaAs~\cite{Jia2015}. 

Quite distinct from the chiral anomaly, the Berry curvature arising from separation of the Weyl nodes leads to other unusual transport effects, particularly the anomalous Hall effect (AHE) and the anomalous Nernst effect (ANE)~\cite{Niu2006, Niu2010}. Unlike conventional system, no ferromagnetism is required for the AHE and ANE in Dirac semimetals because of the strong Berry curvature produced by each of the Weyl nodes. The anomalous Hall conductivity is expressed as~\cite{Ran,Xiaoliang2013}, 
\be
\sigma_{\mathrm{AHE}} = \frac{e^2}{2\pi h} \left|\sum \Delta \bm{k}_i\right|
\label{AHE}
\ee
where $\Delta \bm{k}_i$ is the distance between the $i^{\mathrm{th}}$ pair of Weyl nodes. The thermopower and Nernst effect in Weyl semimetals has been calculated in the Boltzmann equation approach~\cite{Fiete,Tewari2015, Tewari2016, Spivak2016}. 

We report measurements of the thermoelectric tensor $S_{ij}$ of Cd$_3$As$_2$ in two samples (A4, A5) in ``set A'' and two samples (B10, B20) in ``set B'' with the applied thermal gradient $-\nabla T||\bf\hat{x}$ and magnetic field $\bf B ||\bf\hat{z}$ (see Ref.~\cite{Liang2015} for details of the electrical transport measurements in set A and set B samples). 
We obtain $S_{xx}$ and $S_{xy}$ as
\begin{eqnarray}
- S_{xx} &=& E_x/|\nabla T| = -(\rho_{xx}\alpha_{xx} + \rho_{yx}\alpha_{xy}) \label{ex} \\
S_{xy} &=& E_y/|\nabla T| = \rho_{xx}\alpha_{xy} - \rho_{yx}\alpha_{xx},		\label{ey}
\end{eqnarray}
where $\alpha_{ij}$ is the thermoelectric linear response tensor, and $\rho_{ij}$ is the resistivity tensor (see Supplement for the details). 

In Dirac semimetals, the AHE and ANE arise because the Berry curvature $\mathbf{\Omega}(\bf k)$ imparts to the carriers an anomalous velocity ${\bf v}_A = \mathbf{\Omega}(\bf k)\times \hbar{\bf \dot{k}}$, i.e. $\mathbf{\Omega}(\bf k)$ acts like an effective magnetic field in $\bf k$ space ($\bf\dot{k}$ is the rate of change of the wavevector $\bf k$)~\cite{Niu2010}. Previously, the AHE was observed in Cd$_3$As$_2$ as a weak, low-$B$, anomaly in the Hall resistivity $\rho_{yx}$ (Ref.~\cite{Liang2015}). The advantage of the Nernst effect is that it is more sensitive to the anomalous contributions~\cite{Behnia2007,Behnia2011,Behnia2015}. This is because the thermoelectric signals are proportional to the derivative of the conductivities as given by the Mott relation~\cite{Ziman}, viz.,
\be
\alpha_{ij} = {\cal A} \left[\frac{\partial\sigma_{ij}}{\partial \varepsilon}\right]_{\zeta}, \quad 
\left({\cal A} = \frac{\pi^2}{3}\frac{k_B^2T}{e}\right),
\label{Mott}
\ee
where $k_B$ is Boltzmann's constant, $e$ the elemental charge and $\zeta$ the chemical potential. 

In high-mobility semimetals, the conventional Nernst signal rises steeply to a sharp Drude-like peak at the peak field $B_p = 1/\mu$ (where $\mu$ is the mobility), and then decreases towards zero when $B\gg 1/\mu$ (the ``dispersive'' field profile is well illustrated by the curves in Ref. \cite{Liang2013}). By contrast, the ANE signal rises to a maximum value in weak $B$ and remains pinned at this plateau value at large $B$; its profile is step-like. 

\begin{figure*}[t]
\includegraphics[width=15 cm]{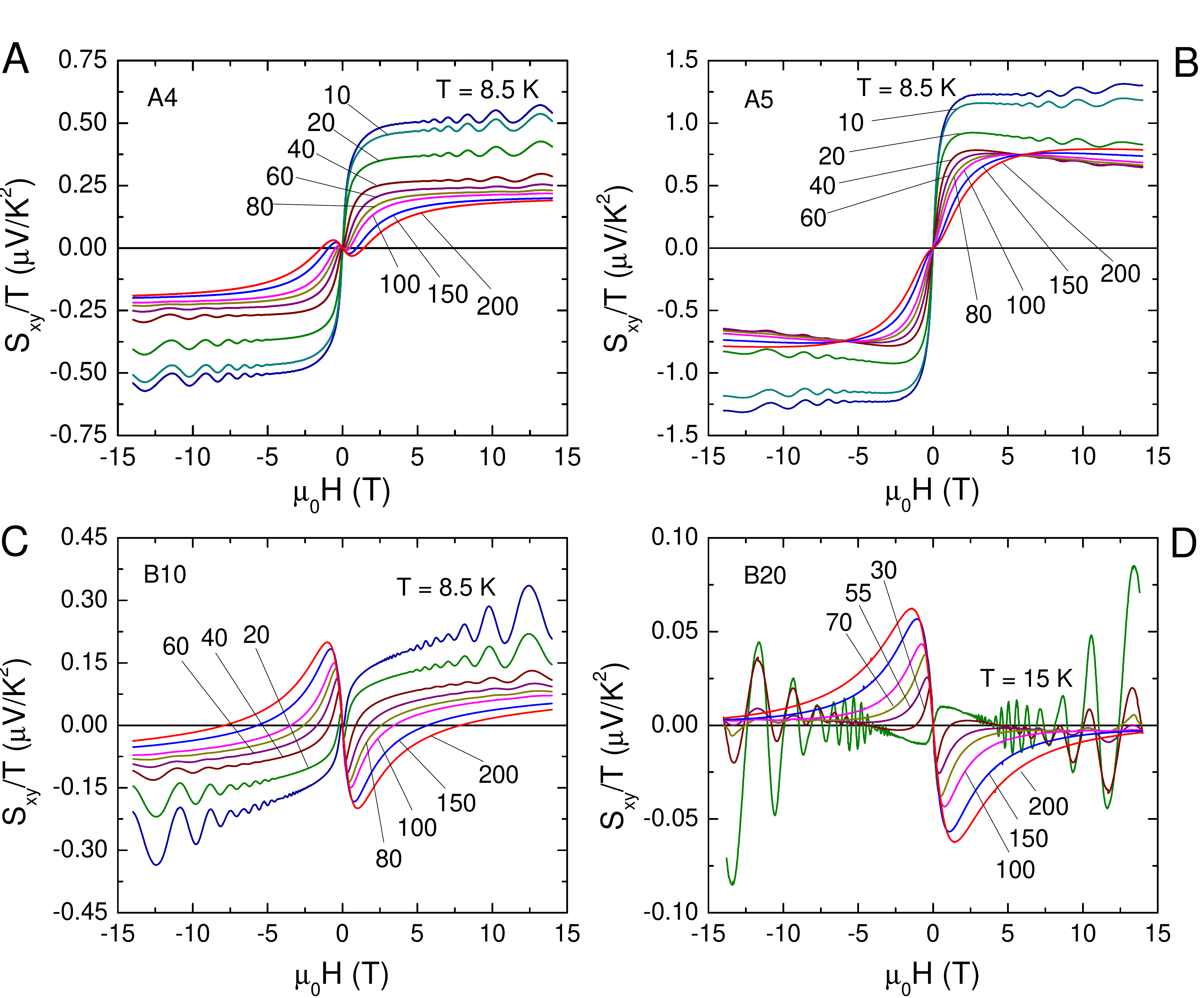}
\caption{\label{Nernst}
(color online) Curves of the Nernst effect at selected temperatures $T$ in samples A4 (Panel A), A5 (Panel B), B10 (Panel C), and B20 (Panel D). In each sample, the ANE component has a step-like profile (especially prominent in Panels A and B), whereas the conventional contribution shows a dispersive (Drude-like) peak (more evident in Panels C and D). In the set A samples A4 and A5, the ANE signal persists to 200 K. By contrast, in set B samples B10 and B20, the conventional Drude-like Nernst signal dominates except at low $T$ ($<$ 30 K) where the ANE dominates.
}
\end{figure*}

Figure~\ref{Nernst} shows the measured Nernst signals at selected temperatures $T$ in samples A4, A5, B10 and B20, respectively. The anomalous component is clearly evident in all 4 samples. The ANE in set A samples dominates the conventional Nernst effect at all $T$ up to 200~K. By contrast, in set B samples, the conventional dispersive profile dominates the signal at high $T$, and the ANE only becomes prominent below 30 K. Theoretically, the separation of the conventional and anomalous contributions to the observed Nernst signal has not been solved in the high-field regime. As an empirical approach, we adopt the following expressions:
\begin{eqnarray}
S_{xy} &=& S_{xy}^N +S_{xy}^A \label{NTot}\\
S_{xy}^N &=& S^N_0 \frac{\mu B}{1+(\mu B)^2}\label{NN}\\
S_{xy}^A &=& \Delta S^A_{xy} \tanh(B/B_0). \label{AN}
\end{eqnarray}
Here, $\mu$ is the carrier mobility, $S^N_0$ is the amplitude of the conventional semiclassical contribution $S_{xy}^N$ (for details, see Ref.~\cite{Liang2013}), $\Delta S^A_{xy}$ is the amplitude of the anomalous Nernst signal $S_{xy}^A$, and $B_0$ is the saturation field above which the signal attains its plateau value $\Delta S^A_{xy}$.

\begin{figure*}[t]
\includegraphics[width=15 cm]{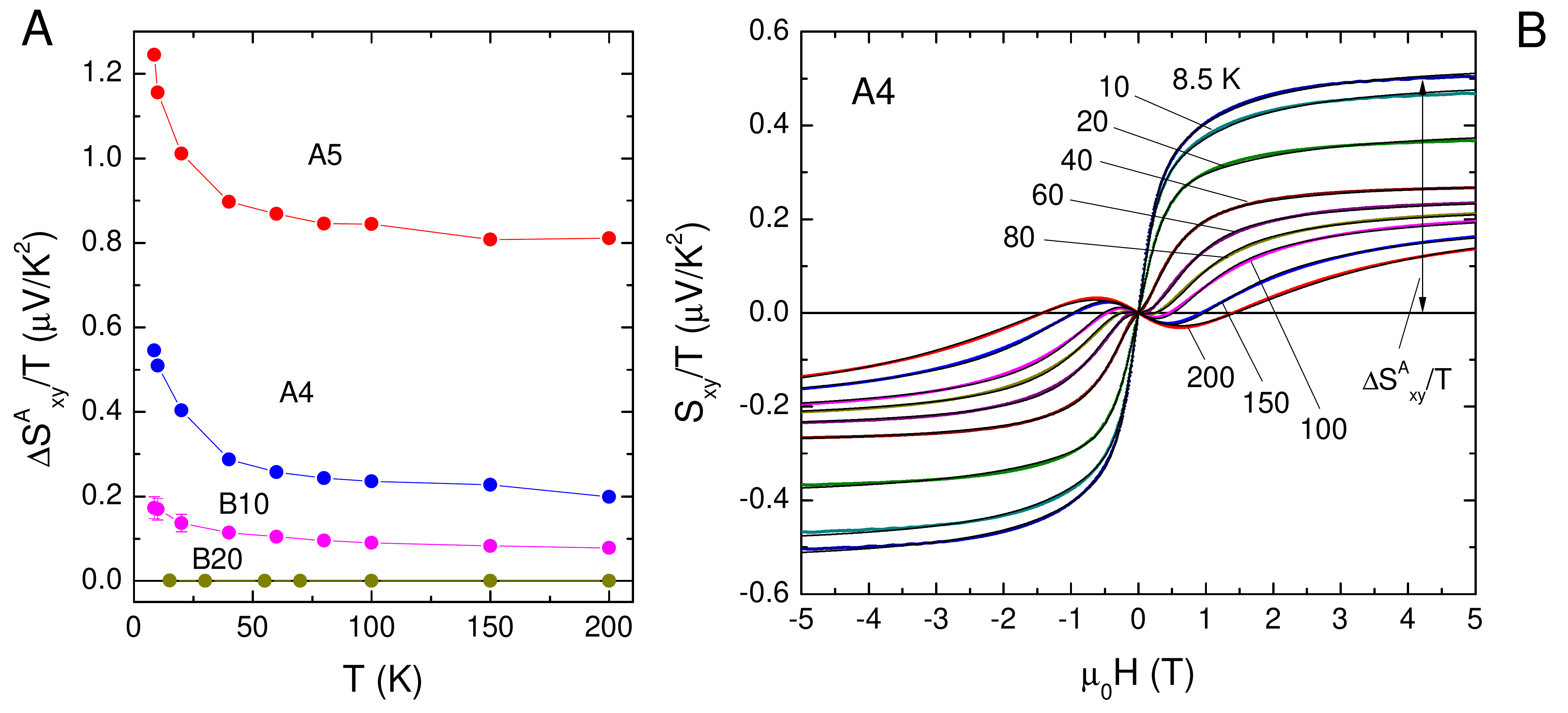}
\caption{\label{NernstAmp} 
(color online). Panel A: The temperature dependence of the amplitude of anomalous Nernst signal $\Delta S^A_{xy}$/T. In the set A samples A4 and A5, the magnitude of anomalous Nernst signals develops rapidly below 50~K, suggestive of a correlation with the mechanism that protects carriers against backscattering.
Panel B: Fits to the observed Nernst effect in sample A4. The observed Nernst effect was fitted to the empirical expression $S_{xy} = S_0^N~\mu B/(1+(\mu B)^2) + \Delta S^A_{xy} \tanh (B/B_0)$, where $S_0^N$ and $\Delta S^A_{xy}$ represent the amplitude of normal and anomalous part of Nernst signals, respectively (Eqs.~\ref{NTot},~\ref{NN},~\ref{AN}). The expression provides good fits (black curves) to the data.
}
\end{figure*}

The empirical expressions provide good fits in all samples. Examples of the fits are shown for sample A4 in Fig.~\ref{NernstAmp}B. The amplitude $\Delta S^A_{xy}$ of the ANE derived from the fits is plotted in Fig.~\ref{NernstAmp}A. Interestingly, while the anomalous Nernst amplitude is small and nearly $T$ independent in set B samples, it is large and strongly $T$-dependent in set A samples. The steep increase below $\sim$ 50 K recalls the $T$ dependence of the transport lifetime $\tau_{tr}$ in set A samples. This suggests a close relation between the ANE and the protection mechanism from backscattering implied by the ultrahigh mobility in set A samples ( $\mu\sim $10$^7$ cm$^2$ V$^{-1}$ s$^{-1}$; see Fig.~1 of Ref.~\cite{Liang2015} and Fig.~\ref{NernstAmp}).

\begin{figure*}[t]
\includegraphics[width=15 cm]{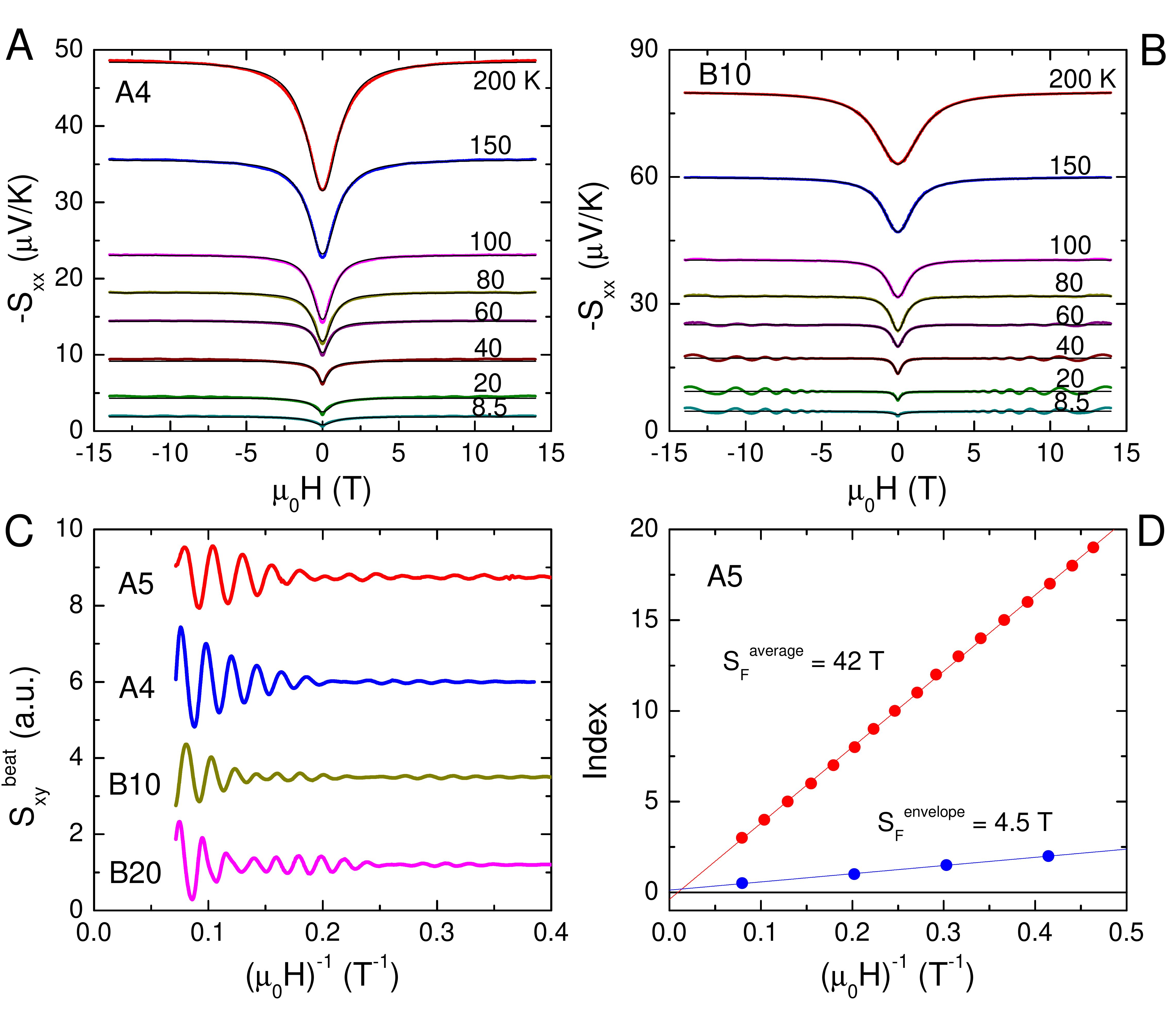}
\caption{\label{Beating}
(color online). Beating pattern in the quantum oscillations in the Nernst signals. Panels A and B show the Seebeck signals of sample A4 (left panel) and B10 (right panel) at selected $T$. The Seebeck signals can be fit using semiclassical expression in Eq. \ref{S}. 
Panel C shows the oscillatory part of the Nernst signals in each sample. Clear beatings were observed for every sample, suggestive of the splitting of Dirac nodes into Weyl nodes. 
Panel D shows the index plot for sample A5 which determines the two Fermi surface areas $S_1$ = 46.5 T and $S_2$ = 37.5 T.
}
\end{figure*}

Next, we discuss the thermopower $S_{xx}$. The measured signals in samples A4, B10 and B20 can be explained by the conventional semiclassical expression (see Ref.~\cite{Liang2013})
\be
S_{xx}(B) = S_0 \frac{1}{1+(\mu B)^2} + S_\infty \frac{(\mu B)^2}{1+(\mu B)^2} \label{S}.
\ee
Here, $S_0$ is the thermopower at $B = 0$ and $S_\infty$ is the limiting value when $B\gg 1/\mu$. For samples A4 and B10, the fits are shown in Panels A, B of Fig.~\ref{Beating}.
(In sample A5, the ultrahigh mobility makes the observed thermopower harder to interpret).

As discussed, the splitting of each Dirac node into two Weyl nodes leads to a finite $\mathbf{\Omega}(\bf k)$. In addition, separation of the Weyl nodes also produces a beating of the bulk quantum oscillations which can be seen in the thermopower and Nernst effect (but are less evident in $\rho_{ij}$ measured on the same samples~\cite{Liang2015}). Panel C of Fig.~\ref{Beating} plots the oscillatory part of the Nernst signals in samples A4, A5, B10 and B20. The beating effect is quite prominent. The macroscopic thickness of the samples (350-1460 $\mu$m) implies that the beating effect arises from interference of closely spaced oscillations in \emph{bulk} states, rather than from surface states related to Fermi arcs. Panel D shows the index plots for the average frequency and the envelope frequency of the beating signal in sample A5. From the slope of the index plot, we extracted the values $S_F^{\mathrm{ave}}$ = 42 T and $S_F^{\mathrm{env}}$ = 4.5 T, from which we obtain two frequencies $S_1$ = 46.5 T and $S_2$ = 37.5 T differing by $\sim 20\%$. Similar values were found for samples A4 ($S_1$ = 50.8 T, $S_2$ = 44.3 T), B10 ($S_1$ = 55.6 T, $S_2$ = 46.9 T), B20 ($S_1$ = 51 T, $S_2$ = 43 T). The beating effect is consistent with the scenario in which the Dirac nodes split into Weyl nodes, leading to distinct Fermi surface cross-section areas.

\begin{figure*}[t]
\includegraphics[width=15 cm]{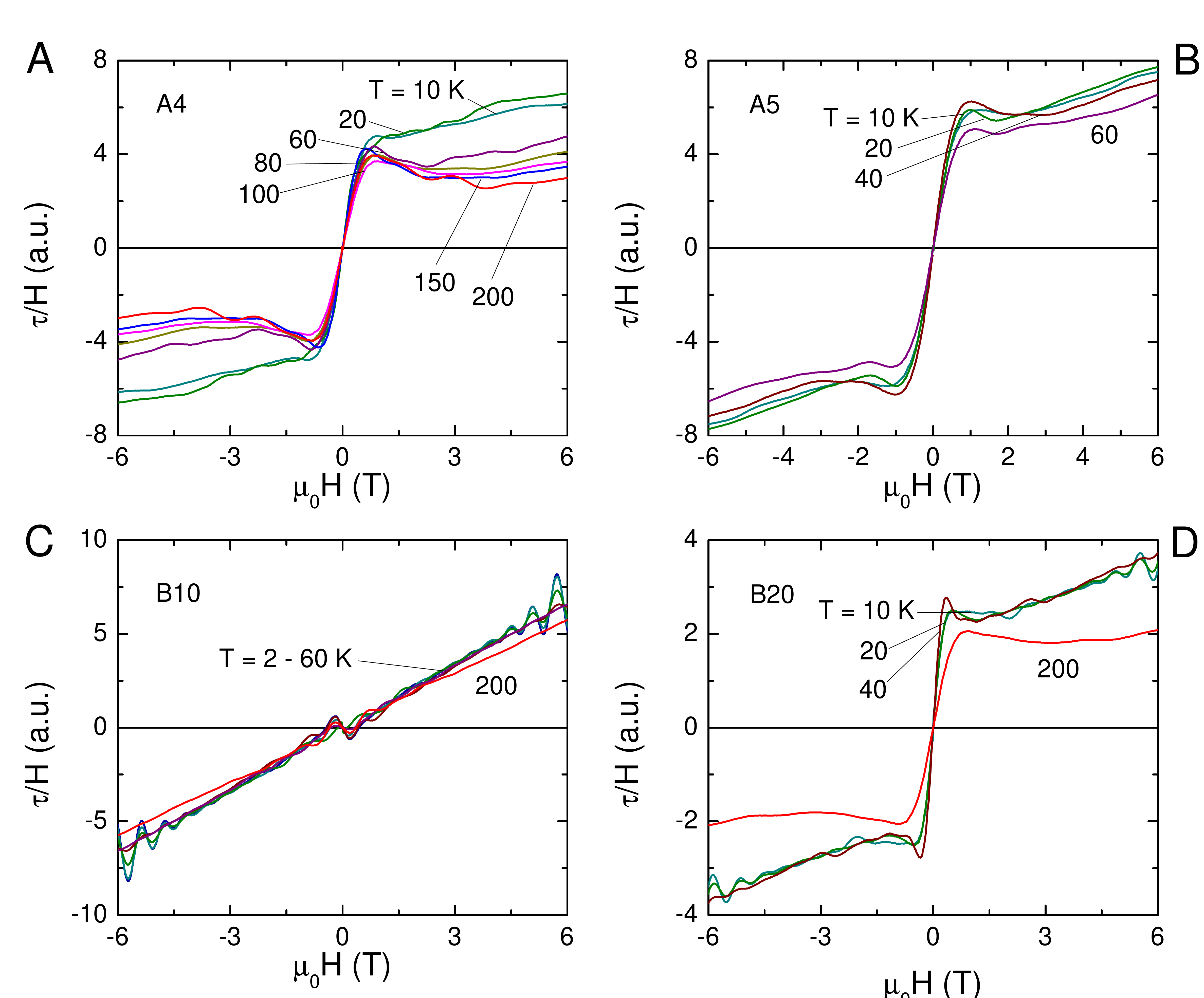}
\caption{\label{M}
(color online). ``Anomalous magnetization'' $M_\tau \equiv \tau/H$ at selected $T$ obtained from torque magnetometry in samples A4, A5, B10 and B20. The anomalous part is confined to $B$ below 1~T in all samples at all $T$ investigated, with a nearly $T$-independent magnitude. These characteristics distinguish $M_\tau$ from the ANE signals in Fig.~\ref{Nernst}, and imply that $M_\tau$ has a different origin from the ANE.
}
\end{figure*}

We also investigated the magnetic response of Cd$_3$As$_2$ via torque magnetometry measurements on samples A4, A5, B10 and B20 (Fig.~\ref{M}). Each of the samples, except for B10, shows an ``anomalous magnetization'' $M_\tau \equiv \tau/H$. This is quite surprising because Cd$_3$As$_2$ does not have magnetic elements. This raises the question whether the observed ANE is related to the ``anomalous magnetization''. At first glance, the anomalous $M_\tau$ is reminiscent of conventional ferromagnetism. However, this scenario is easily excluded by comparing the data of $M_\tau$ taken via torque magnetometry with the magnetization data measured by regular SQUID magnetometry. Both the anomalous $M_\tau$ and the ANE signals are unchanged whether we cool in a finite field or in zero field (see Supplement). By contrast, the step-like magnetization observed in the SQUID data appears only when the sample is cooled in a finite field. To us, it is highly unlikely that the ANE arises from conventional ferromagnetism.

A second question is whether the anomalous $M_\tau$ is coming from the orbital magnetization~\cite{Niu2010,OrbitalMag} generated by $\mathbf{\Omega}(\bf k)$. If this is the case, the ANE and the ``anomalous magnetization'' should show the same dependences on both $B$ and $T$. However, our experiments also exclude this scenario. In all samples, the anomalous $M _\tau$  is restricted to fields well below $\sim 1$~T at all $T$ investigated (its magnitude which persists to 200 K is nearly $T$ independent). By contrast, the magnitudes of the ANE increase rapidly below $\sim 50$~K in set A samples. The onset fields of the anomalous Nernst signals also increase up to $\gtrsim 5$~T at $200$~K, in strong contrast with the behavior of $M_\tau$. Finally, in sample B10, the ANE is finite whereas the anomalous $M_\tau$ signal is absent altogether. Therefore, we conclude that the ANE and the anomalous $M_\tau$ have very different origins (further discussion on this point is given in the Supplement). 

In conclusion, we have performed a detailed investigation of the thermoelectric tensor in Cd$_3$As$_2$ for both set A and set B samples. The Nernst signals reveal a large ANE, suggestive of the existence of Berry curvature $\mathbf{\Omega}(\bf k)$ produced by separation of the Weyl nodes in applied $\bf B$. We also observe a significant beating effect in the quantum oscillations of the Nernst signals. The magnitude of the anomalous part of Nernst signals can be extracted via the phenomenological expressions Eqs.~\ref{NTot},~\ref{NN},~\ref{AN}, whose temperature dependence in set A samples shows a rapid increase below $\sim $ 50 K. The strong increase of $\tau_{tr}$ below 50 K suggests a close relation between the ANE and the mechanism that protects the carriers from backscattering. 


%

\newpage

\vspace{1cm}\noindent

\vspace{1mm}
$^\ast${These authors contributed equally to this work.}

\vspace{3mm}
$^\dagger${Current address of T.L.: Department of Applied Physics, Stanford University, Stanford, CA, 94305}

\vspace{5mm}\noindent
{\bf Acknowledgements} T.L. acknowledges a scholarship from Japan Student Services Organization. N.P.O. acknowledges the support of the Gordon and Betty Moore Foundation's EPiQS Initiative through Grant GBMF4539. R. J. C. and N. P. O. acknowledge support from the U.S. National Science Foundation (MRSEC Grant DMR 1420541). 

\vspace{5mm}\noindent
{\bf Author Information} The authors declare no competing financial interests. Correspondence and requests for data and materials should be addressed to T.L. (liang16@stanford.edu) or N.P.O. (npo@princeton.edu).



\newpage
\clearpage

\renewcommand{\thefigure}{S\arabic{figure}}
\renewcommand{\thesection}{S\arabic{section}}
\renewcommand{\theequation}{S\arabic{equation}}

\setcounter{equation}{0}
\setcounter{figure}{0}
\setcounter{table}{0}

\vspace{4mm}
{\bf Supplementary Information\\
} 
\vspace{6mm} 
	
	\section{Measurements of  S$_{xx}$ and S$_{xy}$}
	In this section, we derive Eqs.~2, 3 used in the main text. The measurements for the thermoelectric tensor $S_{ij}$ were performed with the applied thermal gradient $-\nabla T||\bf\hat{x}$ and magnetic field $\bf B ||\bf\hat{z}$ .In a finite sample, diffusion of carriers down the gradient leads to an opposing electric field $\bf E$ which is detected as the thermopower signal $S_{xx} = -E_x/|\nabla T|$, and the Nernst signal $S_{xy} = E_y/|\nabla T|$. In an infinite medium, the total current density is given by ${\bf J} = \boldsymbol{\sigma}\cdot {\bf E} +  \boldsymbol{\alpha}\cdot(-\nabla T)$~\cite{Ziman}. Here $\sigma_{ij}$ is the conductivity tensor and $\alpha_{ij}$ is the thermoelectric linear response tensor. Setting $\bf J$ = 0 for a finite sample and solving for $\bf E$ (with $\bf B||\hat{z}$ and $-\nabla T||\bf \hat{x}$), we obtain $S_{xx}$ and $S_{xy}$ as
	\begin{eqnarray}
		- S_{xx} &=& E_x/|\nabla T| = -(\rho_{xx}\alpha_{xx} + \rho_{yx}\alpha_{xy}) \label{ex} \\
		S_{xy} &=& E_y/|\nabla T| = \rho_{xx}\alpha_{xy} - \rho_{yx}\alpha_{xx},		\label{ey}
	\end{eqnarray}
	where $\rho_{ij}$ is the resistivity tensor. 
	
	We define the sign of the Nernst signal to be that of the $y$-component of the E-field $E_y$. More generally,
	if ${\bf E}_{\rm N}$ is the E-field produced by the Nernst effect,
	the sign of the Nersnt signal is that of the triple product ${\bf E}_{\rm N}\cdot{\bf B\times}(-\nabla T)$.
	This agrees with Bridgman's ``Amperean current'' convention~\cite{Bridgman1924}, and 
	also with the one adopted for vortex flow in superconductors~\cite{Yayu06}.
	
	\section{``Anomalous Magnetization'' -- Supplemental Data and Discussions}
	
	We discuss more in detail the ``anomalous magnetization'' observed via torque magnetometry measurements in Cd$_3$As$_2$.
	
	\subsection{Exclusion of the Conventional Ferromagnetism}
	
\begin{figure}[h]
		\includegraphics[width=8.5 cm]{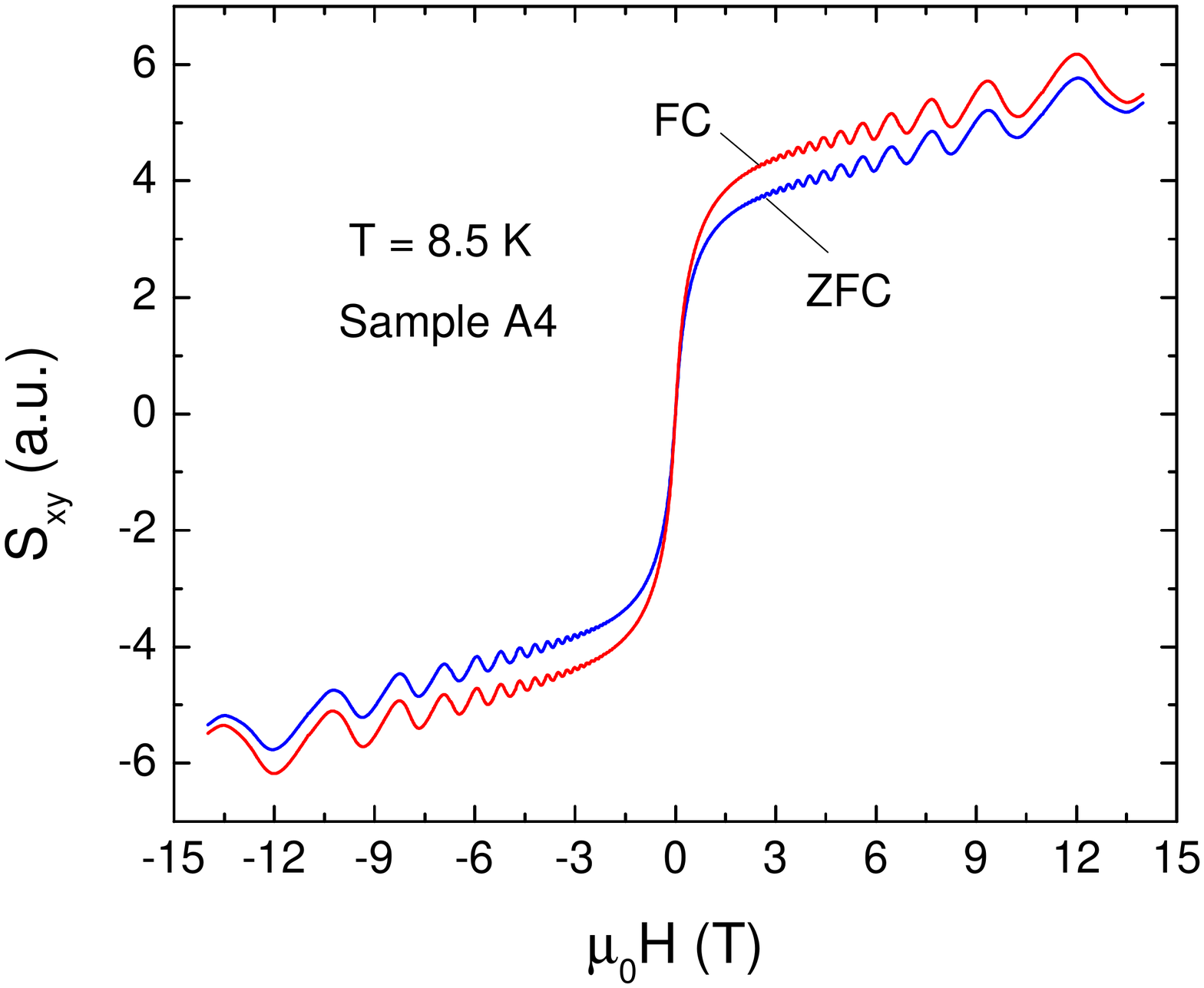}
		\caption{Nernst signals of sample A4 under ZFC and FC. The data are the same within the experimental accuracy, showing that the ANE is inert to the process of ZFC and FC. \label{A4_N}
		}
	\end{figure}

\begin{figure*}[h]
		\includegraphics[width=15 cm]{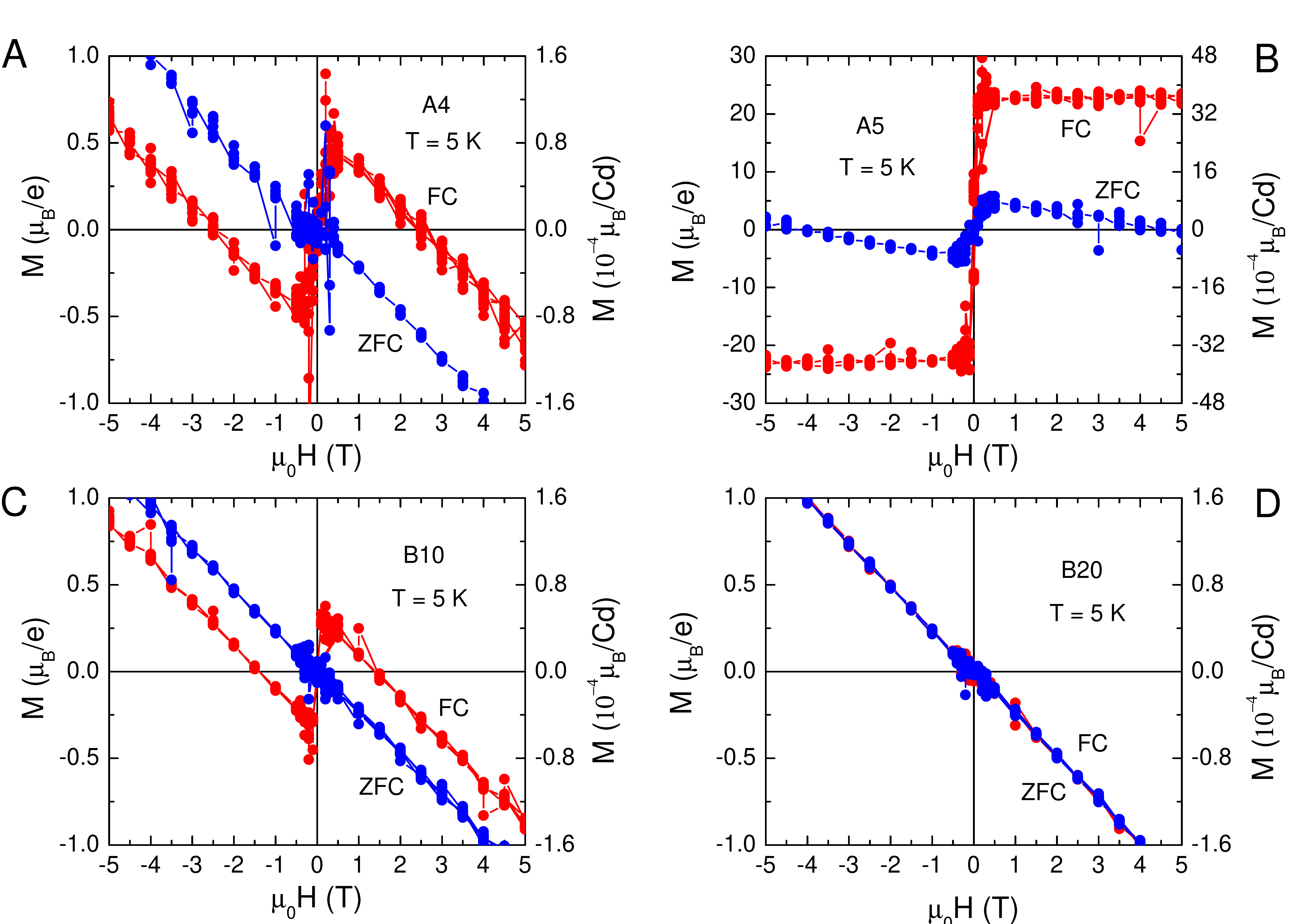}
		\caption{Magnetization for samples A4, A5, B10, B20, obtained from the SQUID measurements. Under ZFC, samples show diamagnetic signals except for sample A5 in which the weak anomaly already enters in. Under FC, every samples except sample B20 show anomaly at low field. The magnitude of the magnetization is plotted for per Cd atom as well as per electron. It ranges $\sim 10^{-4} - 3 \times 10^{-3} \mu_B$/Cd atom and is too large considering the fact that high purity elements with possible impurity below one part in million are used for synthesis. This raises the interesting question of whether the ``anomalous magnetization'' is coming from the orbital magnetization which consists of conducting electrons. For this purpose, we also plotted the magnitude of magnetization per conducting electron, which ranges $\sim 0.5 -25 \mu_B$/e. \label{SQUID}
		}
	\end{figure*}

	As discussed in the main text, Cd$_3$As$_2$ manifests ``anomalous magnetization'' as shown in Fig.~4 of the main text. At first glance, this raises the possibility that the anomalous Nernst effect (ANE) might arise from the conventional ferromagnetism. However, this scenario can be easily excluded by paying attention to the SQUID data shown in Fig.~\ref{SQUID}. As shown in Fig.~\ref{SQUID}, the anomaly only appears for field cooling (FC) data (except for sample A5 which already shows weak anomaly even under zero field cooling (ZFC)). By contrast, the ANE is observed for every samples at ZFC as shown in Fig.~1 of the main text. This sharply excludes the possibility of conventional ferromagnetism playing a role in ANE. Furthermore, as shown in Fig.~\ref{A4_N}, the ANE is inert to the process of FC and ZFC.  We also note that the magnitude of the magnetization taken via SQUID shows that the absolute value is $\sim 10^{-4} - 3 \times 10^{-3} \mu_B$/Cd atom. Considering the fact that high purity elements with possible impurity below one part in million are used for synthesis, this again excludes the possibility of conventional ferromagnetism. We also note that both Cd and As are non-magnetic elements.
	
	\subsection{Orbital Magnetization: Theoretical Backgrounds}
	
	As mentioned in the section above, the ANE does not come from the conventional ferromagnetism. This raises the interesting question of whether or not the ``anomalous magnetization'' arises from the orbital magnetization coming from the Berry curvature $\bm{\Omega}_{\bm{k}}$. Since the origin of orbital magnetization comes from the Berry curvature $\bm{\Omega}_{\bm{k}}$, it has close relation to the AHE and ANE as shown below.
	
	First, we start from a heuristic argument based on Ref.~\cite{OrbitalMag} which relates the thermoelectric component $\alpha_{xy}$ to the orbital magnetization $M_z$, viz.,
	\begin{eqnarray}
		\alpha_{xy} = dM_z/dT \label{Russian}
	\end{eqnarray}
	This simple expression Eq.~\ref{Russian} comes as follows. Suppose we have an uniform system with magnetization $M_z$ pointing along $z$-axis. There exists magnetization current $\nabla \times \bm{M}$. For uniform system, this magnetization current vanishes inside the bulk and only the contribution along the boundary remains. The current $I$ flowing along the edge and the magnetization $M_z$ has a relation $I = M_z L_z$ with $L_z$ the length of the system along $z$-axis. Now, we consider an infinite system along $x$-axis and apply the thermal gradient $-dT/dy$ along $y$-axis and calculate the current density $J_x$ flowing along $x$-axis. Using the above arguments, the total current flowing along the $x$-axis is,
	\begin{eqnarray}
		I_x &=& I(T_0) - I(T_y)\\
		&=& (M(T_0) - M(T_y))L_z\\ 
		&=& -dM/dT ~~ dT/dy ~~ L_yL_z\\
		\therefore J_x &\equiv& \alpha_{xy} ~~ (-dT/dy) =  dM/dT ~~ (-dT/dy)
	\end{eqnarray}      
	yielding the result of Eq.~\ref{Russian}. This suggests that if there is anomalous $\alpha_{xy}$ (hence anomalous Nernst effect), then anomalous orbital magnetization $M$ comes along.
	
	Orbital magnetization can also be related to the anomalous Hall conductivity $\sigma_{xy}$. Here, we follow the pedagogical arguments used in Ref.~\cite{Niu2010} and consider 2D system for simplicity. Suppose we have a potential $V(x)$ that depends on the position $x$ which confines the electrons inside the bulk (see Fig.~8 in Ref.~\cite{Niu2010}). Under Berry curvature $\Omega_z$ along $z$-axis, this produces anomalous velocity $1/\hbar ~ \nabla~ V(\bm{r}) \times \bm{\Omega} (\bm{k})$ along $y$-axis, which is essentially the anomalous Hall component $\sim \sigma_{xy} E_x$. Since the orbital magnetization $M$ is the same as the current $I$ flowing along the boundary,
	\begin{eqnarray}
		M = I &\propto& \int f(\varepsilon + V(x))~\sigma_{xy} (\varepsilon + V(x)) ~E_x~dx\\
		&=& \int f(\varepsilon + V)~\sigma_{xy} (\varepsilon + V) ~dV\\
		&=& \int f(\varepsilon)~\sigma_{xy} (\varepsilon) ~d\varepsilon \label{M_AHE}
	\end{eqnarray}
	with $\sigma_{xy}$ the anomalous Hall conductivity at zero temperature with Fermi energy $\varepsilon$ and $f(\varepsilon)$ the Fermi-Dirac function.
	
	Eqs.~\ref{Russian},~\ref{M_AHE} show that $\sigma_{xy}$ (AHE), $\alpha_{xy}$ (essentially ANE), and orbital magnetization $\bm{M}$ are from the same origin, coming from the Berry curvature $\bm{\Omega}_{\bm{k}}$. Since the Dirac semimetal Cd$_3$As$_2$ has monopoles and anti-monopoles (Weyl nodes) generating the Berry curvature $\bm{\Omega}_{\bm{k}}$, it is natural to ask whether the system also shows orbital magnetization $\bm{M}$ in addition to the AHE and ANE. \\\\
	
	\subsection{``Anomalous Magnetization'' : Possibility of Orbital Magnetization?}
	
\begin{figure*}[h]
		\includegraphics[width=15 cm]{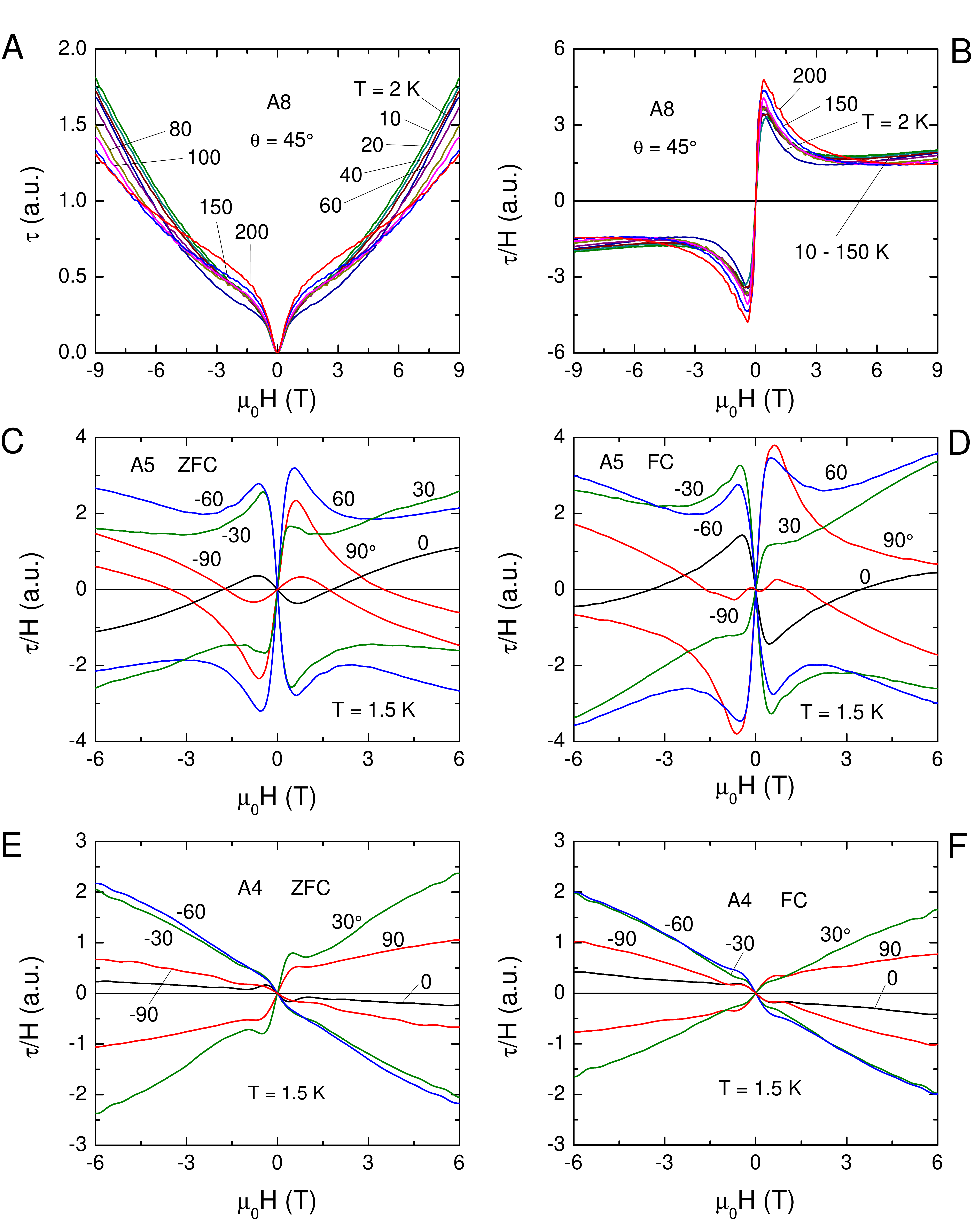}
		\caption{Results of torque magnetometry for samples A8, A5 and A4. Panel A shows the torque data for sample A8 at selected temperatures up to 200~K. At low fields below 1~T, cusps are observed, corresponding to the ``anomalous magnetization'' $M_\tau = \tau/H$ shown in panel B. Panels C, D (E, F) show the angular dependence of the ``anomalous magnetization'' obtained for sample A5 (A4) for ZFC (panels C, E) and FC (panels D, F), showing that no significant differences appear between the two procedures. \label{A8_A5_A4}
		}
	\end{figure*}

	As previous section implies, orbital magnetization, AHE, ANE have close relation, sharing the same origin of Berry curvature $\bm{\Omega}_{\bm{k}}$. If the ``anomalous magnetization'' is the manifestation of orbital magnetization, it should show the same trend of temperature and field dependence as the ANE. To test whether the ``anomalous magnetization'' is coming from the orbital magnetization or not, we performed detailed measurements of torque magnetometry and investigated in detail the temperature, magnetic field, angle, and sample dependences as shown in Fig.~4 of main text and Fig.~\ref{A8_A5_A4}. 
	
	Fig.~\ref{A8_A5_A4} shows the torque signals $\tau$ for samples A8, A5 and A4. Panel A shows the torque curves at selected temperatures plotted as a function of applied magnetic field. At low fields, there are cusps corresponding to the ``anomalous magnetization'' as shown in panel B which plots $M_\tau \equiv \tau/H$. The curves are similar at all temperatures investigated, with the anomaly confined below $\sim 1$~T. The comparison of the ZFC and FC is plotted for sample A5 (A4) in panels C, D (E, F), which show similar angular dependence and no significant difference was observed between the two procedures. 
	
	While the scenario of ``anomalous magnetization'' arising from orbital magnetization and coming from the Berry curvature $\bm{\Omega}_{\bm{k}}$ is interesting and attracting, the data of Fig.~4 of main text and Fig.~\ref{A8_A5_A4} suggest the different origins between the ``anomalous magnetization'' and the ANE. First, the anomaly of ``anomalous magnetization'' is confined below $\sim 1$~T for every angle, temperature, and samples (we note that tilting the angle in the torque magnetometry measurements could only have the effect of broadening the anomaly and the true anomaly is confined well below 1~T). By contrast, the ANE only onsets above $\sim 5$~T at 200~K. This shows sharp contrast between ``anomalous magnetization'' and ANE, implying the different origins between them~\footnote{The arguments here regarding the onset field of the anomaly can be used to exclude the possibility of the conventional ferromagnetism as well.}. Furthermore, while the profile of ``anomalous magnetization'' remains essentially the same all the way up to $\sim 200$~K, ANE for set A samples rapidly develops below $\sim 50$~K, again suggesting the different origins between them. We also note that for sample B10, while the ``anomalous magnetization'' vanishes, the ANE is finite. 
	These results suggest the different origins between the ``anomalous magnetization'' and the ANE. Investigation of the origin of ``anomalous magnetization'' is a fruitful direction to pursue in the future.
	
	
	%



\vspace{1mm}
$^\ast${These authors contributed equally to this work.}

\vspace{3mm}
$^\dagger${Current address of T.L.: Department of Applied Physics, Stanford University, Stanford, CA, 94305}

\vspace{5mm}\noindent
{\bf Author Information} The authors declare no competing financial interests. Correspondence and requests for data and materials should be addressed to T.L. (liang16@stanford.edu) or N.P.O. (npo@princeton.edu).

\end{document}